\documentclass[reprint, prb, amsmath, amssymb, superscriptaddress]{revtex4-1}

\usepackage[T1]{fontenc} 
\usepackage{xcolor}
\usepackage{hyperref}
\usepackage{graphics}
\usepackage{gensymb}

\begin{document}

\title{Scanned single-electron probe inside a silicon electronic device}
\author{Kevin S. H. Ng}
\affiliation{Centre for Quantum Computation and Communication Technology, School of Physics, The University of New South Wales, Sydney, New South Wales 2052, Australia}
\affiliation{5. Physikalisches Institut and Center for Integrated Quantum Science and Technology, Universit{\"a}t Stuttgart, Pfaffenwaldring 57, 70569 Stuttgart, Germany}
\author{Benoit Voisin}
\affiliation{Centre for Quantum Computation and Communication Technology, School of Physics, The University of New South Wales, Sydney, New South Wales 2052, Australia}
\author{Brett C. Johnson}
\affiliation{Centre for Quantum Computation and Communication Technology, School of Physics,
University of Melbourne, Melbourne, Victoria 3010, Australia}
\author{Jeffrey C. McCallum}
\affiliation{Centre for Quantum Computation and Communication Technology, School of Physics,
University of Melbourne, Melbourne, Victoria 3010, Australia}
\author{Joe Salfi}
\affiliation{Centre for Quantum Computation and Communication Technology, School of Physics, The University of New South Wales, Sydney, New South Wales 2052, Australia}
\affiliation{Department of Electrical and Computer Engineering, University of British Columbia, Vancouver, BC V6T 1Z4, Canada}
\email{jsalfi@ece.ubc.ca}
\author{Sven Rogge}
\affiliation{Centre for Quantum Computation and Communication Technology, School of Physics, The University of New South Wales, Sydney, New South Wales 2052, Australia}



\graphicspath{{./figures/}}



\begin{abstract}
 
Solid-state devices can be fabricated at the atomic scale, with applications ranging from classical logic to current standards and quantum technologies. While it is very desirable to probe these devices and the quantum states they host at the atomic scale, typical methods rely on long-ranged capacitive interactions, making this difficult. Here we probe a silicon electronic device at the atomic scale using a localized electronic quantum dot induced directly within the device at a desired location, using the biased tip of a low-temperature scanning tunneling microscope. We demonstrate control over short-ranged tunnel coupling interactions of the quantum dot with the device's source reservoir using sub-nm position control of the tip, and the quantum dot energy level using a voltage applied to the device's gate reservoir. Despite the $\sim 1$~nm proximity of the quantum dot to the metallic tip, we find the gate provides sufficient capacitance to enable a high degree of electric control. Combined with atomic scale imaging, we use the quantum dot to probe applied electric fields and charge in individual defects in the device. This capability is expected to aid in the understanding of atomic-scale devices and the quantum states realized in them.

\end{abstract}

\maketitle




The miniaturization of solid-state devices has driven tremendous improvements in their performance and functionality. For example, the information revolution has been enabled through decades of continuous miniaturization of silicon complementary metal-oxide-semiconductor devices. Device miniaturization also permits the localization and control of single electrons in solids, which has led to new fundamental physics experiments\citep{Goldhaber1998}, new current standards \citep{Pekola2013, Rossi2014} and could enable the realization of quantum computers \citep{Loss1998, Kane1998, Petta2005,Koppens2006} and quantum simulators\citep{Feynman1982, Singha2011, Salfi2016, Hensgens2017}. Such single electron devices can employ quantum dots (QD) \citep{Kouwenhoven2001, Hanson2007, Zwanenburg2013} or active atoms placed with atomic-scale precision in solids using scanning tunnelling microscopy (STM)\citep{Fuechsle2012, Zwanenburg2013, Folsch2014, Huff2018, He2019}.   

With ever-shrinking device sizes, real-space techniques for probing their properties and the complex states they host are becoming increasingly important. One notable technique, the scanning single electron transistor\citep{Yoo1997} (SSET) monitors the capacitive response of an electron localized on scanning single-electron transistor to probe electrons in a device. Another notable technique, scanning gate microscopy (SGM)\citep{Sellier2011} records the electrostatic response of the device's conductance to a scanned conductive tip. The  interaction mechanism at play in the SSET and SGM is capacitive (electrostatic), which inherently limits functionality and spatial resolution.  

Here we go beyond the electrostatic interaction paradigm of SSET and SGM by showing that a spatially localized single-electron QD probe can be induced at an arbitrary location inside an electronic device, and made to interact not only capacitively, but also through tunneling interactions with electrons in the device. The method relies on electrostatically inducing a QD at a desired location with sub-nm precision using a biased atomically sharp tip of a low temperature scanning tunnelling microscope (LT-STM) operating at 4.2 Kelvin\citep{Dombrowski1999, Freitag2016, Salfi2018}. Our device consists of donor implanted source and gate reservoirs and the QD probe is induced within the device, beneath a hydrogen terminated silicon surface. This surface is atomically flat, similar to surfaces found in other materials relevant for electronic devices including two-dimensional materials\citep{Dombrowski1999, Freitag2016}. 

\begin{figure*}
\includegraphics{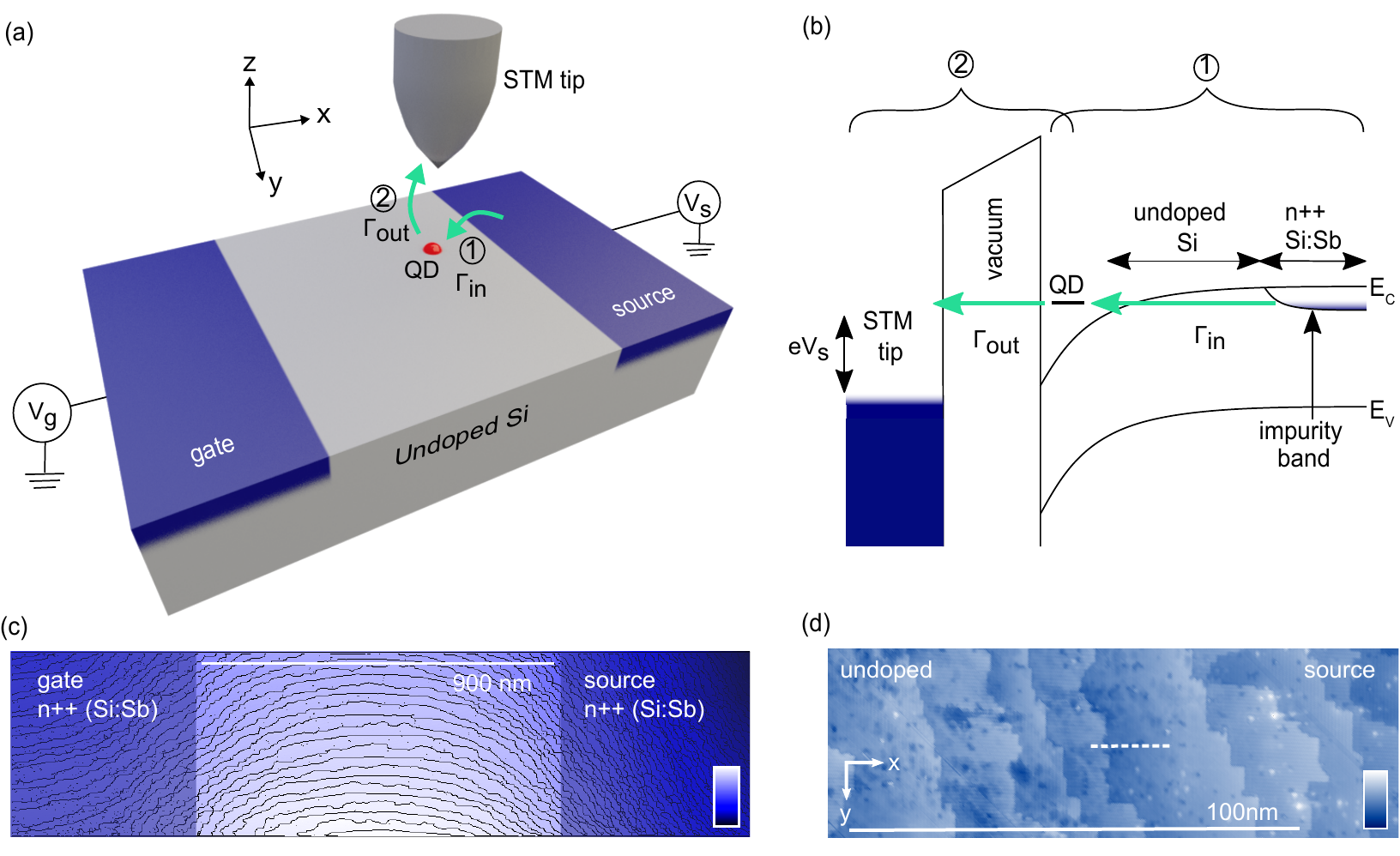}
		\caption[schematic_experiment]
		{Induced QD probe. (a) Experimental schematic of the device, where single electron tunnelling occurs from a biased antimony donor (source) reservoir to the QD and out to the tip. By moving the tip closer to the source reservoir, the tunnel coupling between QD and reservoir is controllably increased. Opposite the source reservoir, an additional antimony reservoir is biased to provide a gate voltage, influencing the energy of the QD. (b) Energy band diagram during resonant electron tunnelling. Single electrons tunnel through a state of the QD created by tip-induced band bending from an applied bias $V_{s}$. $E_C$ and $E_V$ are the conduction and valence band edges, respectively. (c) Global topography of the device (nominal shown). Step edges, identified by an edge finding algorithm and superimposed on the data, are found to be bunched in the vicinity of the gate/undoped and undoped/source junctions. Scale: $0$ -- $7.2$~nm. (d) STM image of the sample surface at the undoped-source junction taken at $V_{s}=-1.6$~V. The junction is identified by a change in the appearance of the dangling bonds. The line indicates the position where the QD measurements presented in Figure~\ref{fig2} are performed. Scale: $0$ -- $0.66$~nm.}\label{fig1}
\end{figure*}

Working in the regime of single-electron tunnelling through the induced QD, we demonstrate the ability to tune the tunnel coupling of the QD to the dopant-atom source reservoir by moving the STM tip, and use this to characterize the decay length of our probe QD wavefunction, which we find to be $\sim 9$~nm. This technique works alongside the traditional use of STM to image device surfaces with atomic resolution. Our study is performed on a multi-terminal device, having both source and gate electrodes (Figure~\ref{fig1}a), which differs significantly from previous studies\citep{Dombrowski1999,Freitag2016,Shim2019, Wagner2019}. We find that the QD state energy can be controlled using the device's gate voltage. The gate lever arm, which expresses the ratio of the capacitive coupling between the QD and the gate, to the total capacitance, is found to be $\sim 0.08$, offering a high degree of electric control.  Surprisingly, the QD capacitance is dominated by the capacitive coupling to the source terminal in the device, despite the $\sim 1$~nm proximity of the QD state to the metallic tip that induces it. 

The QD state energy also reacts to atomic scale defects that we directly observe with atomic resolution, allowing us to map the defect charge state directly using the QD. Extended to devices based on arrays of dopant atoms placed with atomic precision, we envision these new experimental capabilities could be used to characterize or enhance the functionality of atomic scale devices. For example, the induced QD could be used to induce a highly tunable super-exchange\citep{Malinowski2019} between donor atoms\citep{Srinivasa2015}, or to implement local spin readout\citep{Elzerman2004} on atomic structures consisting of interacting dopant atoms forming quantum simulators\citep{Salfi2016}. 



Before discussing our experiment, we first describe the concept of our induced QD probe. When a negative sample bias $V_{s}$ is applied to a semiconductor with respect to an STM tip, an electric field is induced by the tip that locally bends the semiconductor bands downward to create an attractive potential for electrons. We use this capability to trap individual electrons beneath our atomically flat surface\citep{Salfi2018}. This is done in the insulating p-type region of a device, fabricated as a planar n$^{++}$/p/n$^{++}$ junction over a p-type substrate (see supp. mat.), where the antimony doped n$^{++}$ regions form electron reservoirs that act as a source and gate (Figure~\ref{fig1}a,c). Since the electric field is strongest below the tip, the QD follows the position of the tip, and in our experiment, we move the QD toward the source reservoir, performing resonant single-electron tunnelling spectroscopy on the QD. Resonant tunnelling (Figure~\ref{fig1}b) is detected as a measured step (peak) in current (conductance) when the QD state is on resonance with the source reservoir. We control the electron tunnel-in rate $\Gamma_{\rm in}$ from the source reservoir to the QD using the sub-nanometre positioning precision of the STM tip along $x$ and $z$ (Figure~\ref{fig1}a) and track the variation of $\Gamma_{\rm in}$ through the measured single-electron tunnelling current $I = e(\Gamma_{\rm in}^{-1}+\Gamma_{\rm out}^{-1})^{-1}$, where $\Gamma_{\rm out}$ is the tunnel rate from the QD out to the tip\citep{Bonet2002} and $e$ is the electronic charge. To observe the variation of $\Gamma_{\rm in}$, we work in the regime where $\Gamma_{\rm in} \ll \Gamma_{\rm out}$, which can be achieved by varying $\Gamma_{\rm out}$ using the tip-sample separation (vacuum barrier width). A voltage $V_{g}$ is applied to the gate reservoir located approximately $900$~nm away from the source reservoir, which we use to tune the energy of the QD. As expected for a single-electron device, this changes the corresponding bias $V_{s}$ at which a QD state is on resonance with the source reservoir.  

\begin{figure*}
\includegraphics{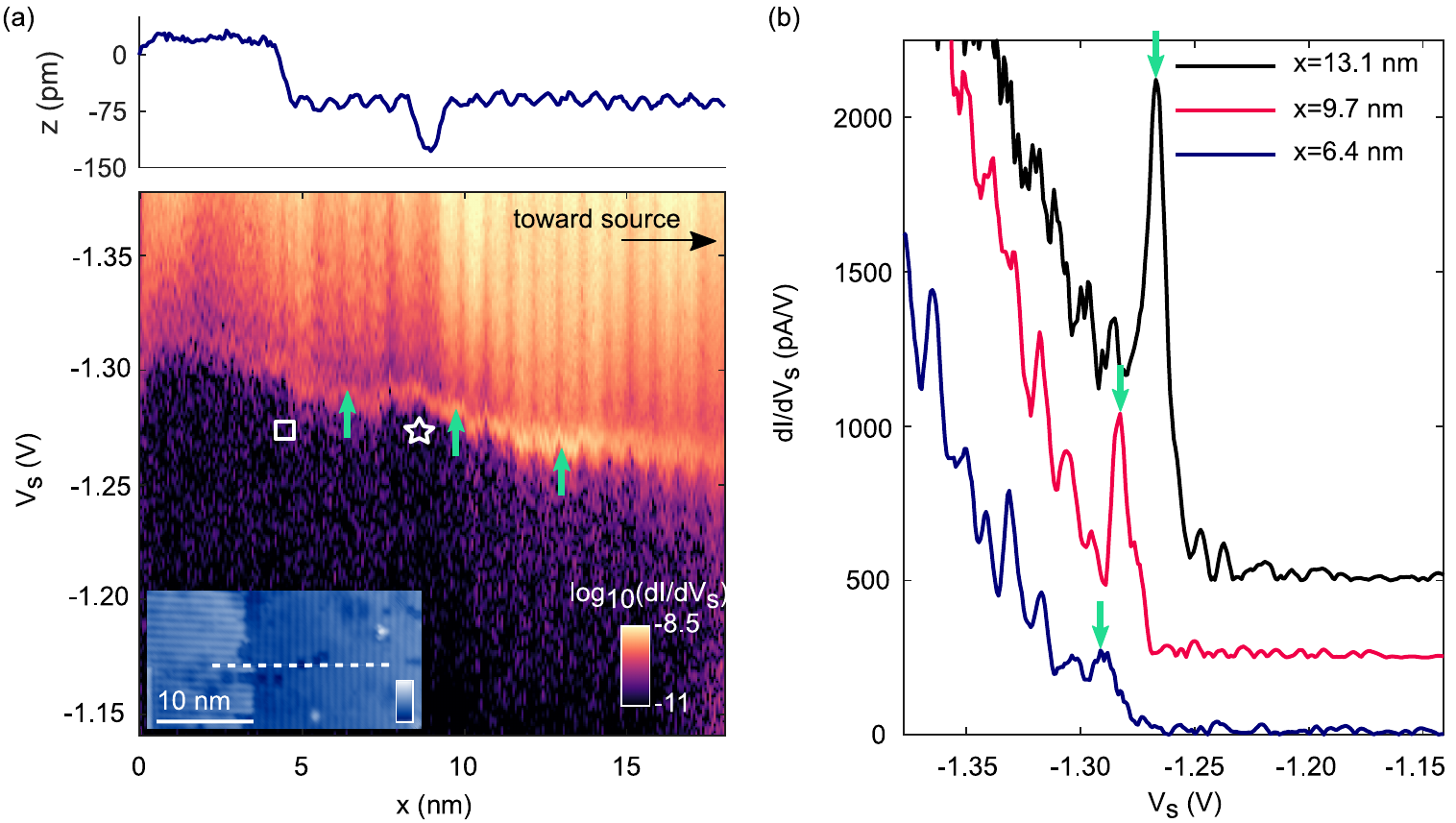}
		\caption[STM_experiment]
		{Spatially resolved resonant single electron tunnelling through the QD. (a) Conductance map taken over an $18$~nm distance at the line and the corresponding surface topography (top). A spatially continuous peak in conductance is observed as the tip is moved toward the reservoir, indicative of single electron tunnelling through a tip induced QD. The positions of the square and star correspond to a terrace step edge and dimer vacancy respectively the QD encountered during measurement. The position of the arrows on the resonance correspond to the traces shown in (b). Taken at $V_g$ = $-1.6$~V. Inset: zoomed-in topography image showing the same line as Figure~\ref{fig1}d, taken at $V_s$ = $-1.6$~V. Scale: $0$ -- $0.38$~nm. (b) Single conductance traces of the conductance map taken at equal intervals along the resonance. Traces are offset by $250$~pA/V for clarity. The height of the single resonance peak indicated by the arrows increases as the tip approaches the reservoir.}\label{fig2}
\end{figure*}



Following fabrication of the device (Figure~\ref{fig1}a and supp. mat.), we first imaged the device at the level of its global topographical features, and then in detail at the junction between the undoped region and the source region (Figure~\ref{fig1}c,d). The device is identified by the bunching of step edges at both the gate/undoped and undoped/source junctions (Figure~\ref{fig1}c). The distinct difference in appearance of dangling bonds in the doped and undoped regions confirms that the step-edge bunching occurs at the junction (Figure~\ref{fig1}d). Dangling bonds appear `bright' (negatively charged) in doped regions and `dark' (positively charged) in undoped regions in filled state imaging\citep{Labidi2015}. Spatially resolved spectroscopy was performed at several lines traversing the junction, such as the one in Figure~\ref{fig1}d.

 

The differential conductance $dI/dV_{s}$, obtained by numerically differentiating the current $I$, is shown as a function of source-tip bias $V_s$ and tip position $x$ in Figure~\ref{fig2}a for the line shown in the STM inset. The corresponding topography $z$ recorded by the tip is plotted above the conductance map. From $x = 0$ -- $4$~nm, no clear resonance is seen as tunnelling to any localized state below the tip is prevented by the wide substrate tunnel barrier. As the tip moves closer to the source reservoir, the resonance peak becomes apparent for $x\gtrsim 4$~nm at a voltage $V_s \approx -1.3$~V, and remains visible for increasing $x$. We also note a weak background corresponding to direct tunnelling from the occupied valence band to the tip energy starting around $V_s\approx-1.1$~V to $-1.2$~V, as expected near an n-type lead \citep{Salfi2014}. Importantly, only the feature emerging at $V_s\approx-1.3$~V has a resonant peak lineshape indicative of a QD state. We plot in Figure~\ref{fig2}b the numerically differentiated conductance for traces of Figure~\ref{fig2}a at the coordinates $x=6.4$~nm, $x=9.7$~nm and $x=13.1$~nm.

The combination of the resonance lineshape, its continuous spatial nature, and its increasing amplitude with decreasing distance from the reservoir allows us to confidently rule out that the resonance is due to tunnelling through stray dopants at the junction or surface dangling bonds\citep{Salfi2014, Taucer2014, Labidi2015}. The possibility that the peak emerging at $x\approx4$~nm consists of a QD state hybridized with dopants or dangling bonds can also be ruled out, because we do not observe the spectroscopic signature of dopants\citep{Salfi2014} and dangling bonds\citep{Taucer2014} in Figure~\ref{fig2}a. For $x\gtrsim 13$~nm however, we note the conductance of the resonance levels off and even slightly decreases. It is difficult to isolate the reason for this decrease but we note that it coincides with the QD reaching localized states\citep{Salfi2014} observed in our measured spectra (supp. mat. Figure~\ref{supp2}). Indeed, fixed localized states in the sample are expected to be found when the tip enters the reservoir region for $x\gtrsim 18$~nm, since the reservoir is composed of antimony donor levels that are energetically below the induced QD resonance. Tunnelling here is expected to be more complex because of the potential hybridization of the induced QD with donors\citep{Salfi2018} which have a high concentration in the reservoir.


\begin{figure}
\includegraphics{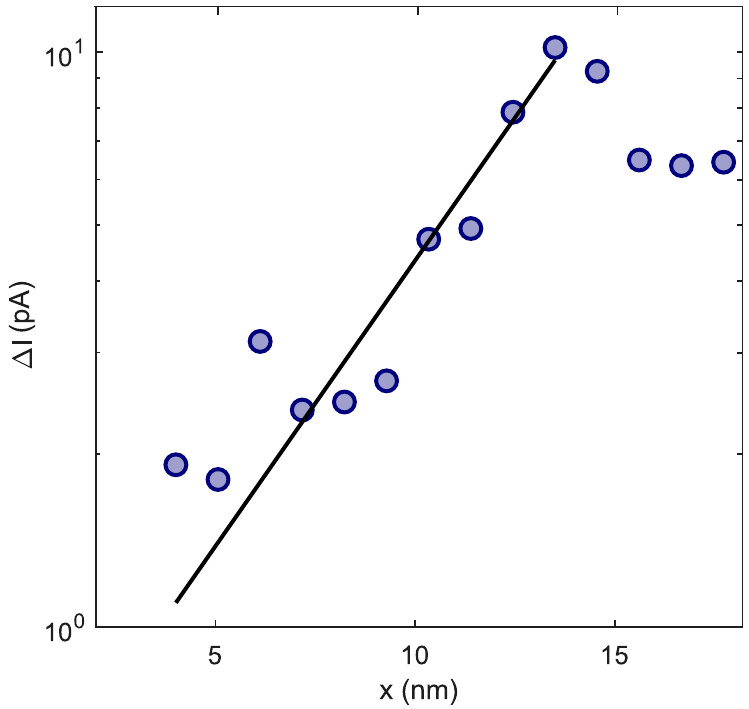}
		\caption[deltaI]
		{Probing the QD wavefunction in space. Change in the induced current during resonant tunnelling as a function of tip position. Prior to the localized states beyond $x$ = $13$~nm, the increase in the current between $x$ = $4.0$ -- $13$~nm due to QD-reservoir wavefunction overlap increases exponentially, where $\kappa = 0.23 \pm 0.06$~nm$^{-1}$ for a fit $\Delta I \propto$ exp($\kappa x$). This gives an estimate of the QD wavefunction decay length $\lambda$ to be around $9$~nm. 
}\label{fig3}
\end{figure}

The QD energy and tunnelling rate to the QD are expected to be influenced when the QD encounters and interacts with surface defects. This is seen in Figure~\ref{fig2}a as two disturbances of the resonance energy in $V_s$, indicated by the square and star that occur at positions where a step-edge and atomic defect are found on the surface, respectively. In measurements performed along a different line towards the reservoir (supp. mat. Figure~\ref{supp1}), we observe strong disturbances of the tunnelling current as the QD encounters accumulated negative charge at a step edge, and from a negatively charged (2e$^{-}$) dangling bond, at $x$ = $24$~nm and $32$~nm respectively. For both disturbances, the bias $V_{s}$ required to maintain resonant tunnelling through the QD decreases (more negative) as it moves toward the negative charge, before subsequently increasing as it moves away. This is expected, because the induced QD energy level is sensitive to surface charge in the device through the Coulomb interaction. 

\begin{figure}
\includegraphics{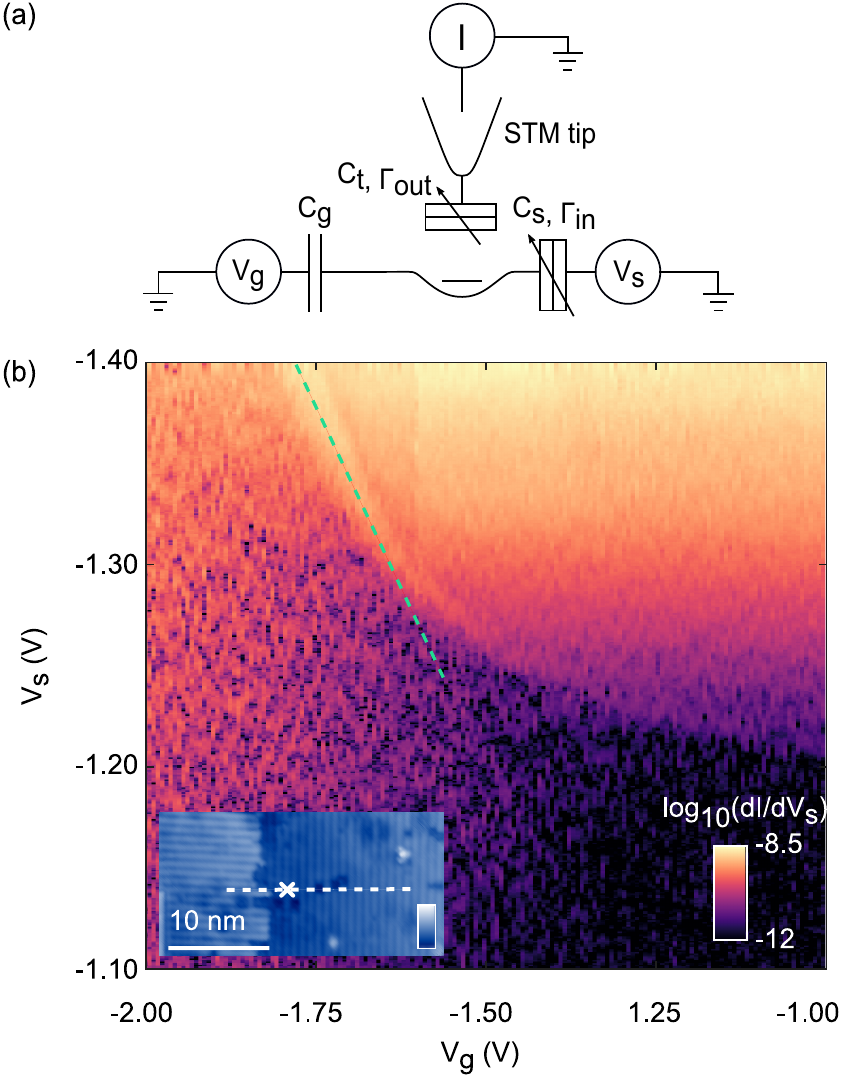}
		\caption[gate_effect]
		{Tuning the QD energy level with the gate. (a) Circuit schematic of the device.  The gate voltage $V_{g}$ is swept with the tip held at the position of the marker in the STM inset of (b). $C_{t}$, $C_{s}$ and $C_{g}$ are the capacitance between the QD and tip, source and gate respectively. (b) Charge stability diagram showing the ability to influence the energy of the QD state with a gate. The resonance follows the green line for values less than $V_g = -1.6$~V, indicating a constant capacitive coupling between the QD and its environment in this voltage range. The non-linear, weaker behaviour of the gate for values greater than $-1.6$~V can be explained by an increase in the stray capacitance experienced by the QD near the source due to the increasingly attractive gate potential. Inset: zoomed-in topography image showing the same line as Figure~\ref{fig1}d, where the marker on the line indicates the position of the tip during measurement. Taken at $V_s$ = $-1.6$~V. Scale: $0$ -- $0.38$~nm.}\label{fig4}
\end{figure}

We now investigate the interaction of the induced QD with the planar electric field established between the biased source and gate reservoirs. In Figure~\ref{fig2}a, the voltage $V_{s}$ required to bring the QD on resonance with the source reservoir increases from $-1.3$~V to $-1.27$~V over a $9.2$~nm distance between $x$ = $4.0$ -- $13.2$~nm as the QD approaches the reservoir. This is expected because of the $x$ oriented electric field $E_{x}$ that exists between the source $V_s$ and the fixed gate $V_g=-1.6$~V during spectroscopy, which introduces an $x$-dependent offset to the QD energy. Assuming a resonance around $V_s=-1.3$~V, a gate voltage $V_g=-1.6$~V, and a source to gate distance of approximately $900$~nm, we estimate  $E_x\approx 0.4$ MV/m, varying slightly for $V_{s}=-1.3$~V to $V_s=-1.27$~V where the resonance is found. Since the gate voltage $V_{g}$ is always more negative compared to $V_{s}$ during measurements, the electric potential is more repulsive when the QD is closer to the gate reservoir. As a result, when the QD moves away from the gate i.e. tip position $x$ increases, less tip-induced band bending and a higher $V_{s}$ is required to bring the QD state on resonance with the source reservoir. From $E_{x}$, we quantify the ability of the sample-tip bias voltage to change the energy of the QD state $\Delta E$ with the lever arm parameter $\alpha_0$, where \(\Delta E = e\alpha_0(V_{s,2}-V_{s,1})\) and $V_{s,1}$, $V_{s,2}$ are differing values of source-tip bias. By equating the energetic change of the resonance in Figure~\ref{fig2}a toward lower values as $x$ increases, to the expected energy level change due to $E_{x}$, we obtain a value of $\alpha_0 \approx 0.1$. This is in good agreement with values of $\alpha_0$ independently obtained by fitting to a model of single electron tunnelling between Fermi reservoirs through a quantized state \citep{Foxman1993}, and in agreement with previously extracted values of the lever arm for a tip-induced QD\citep{Salfi2018}. Using this model, where fitting is facilitated by increased smoothing of the measured current, we find $\alpha_0$ varying from $\approx$ $0.14$ to $0.08$ for $x$ increasing between $11.8$ -- $17.8$~nm, as the induced QD approaches the reservoir. For the two different intervals in $x$ which only have approximately $1$~nm overlap, the estimated values for $\alpha$ match closely, giving us confidence in the obtained values. We note that varying $V_{s}$ and $x$ could also change the QD potential and its energy spectrum\citep{Dombrowski1999}, however, the behaviour of the ground state seems to follow a relatively simple model with an energy shift of the level due to the $x$ oriented electric field.

We estimate the decay of the tail of the induced QD wavefunction in space using the measured change in induced current $\Delta I(x)$ through the QD as it is moved toward the source reservoir with increasing $x$. We plot in Figure~\ref{fig3} $\Delta I(x)$ using the measured current of Figure~\ref{supp2} (supp. mat.) before the QD encounters the localized states. The blue dots plotted on the resonance of the conductance map indicate the position of the traces in $x$ and the peaks in $V_{s}$ used. For $14$ traces along the resonance, we find $\Delta I(x)$ exponentially increasing between $x$ = $4.0$ -- $13$~nm before slightly decreasing $\approx$ $5$~nm prior to encountering the localized states. This observation suggests that the appearance of the resonance at $x\approx 4$~nm is due to the increasing tunnel coupling $\Gamma_{\rm in}(x)$ as the QD moves towards the source reservoir. Indeed, we work at a tip-sample separation that establishes a tunnel rate where $\Gamma_{\rm out}(z)$ is larger than the tunnel rate from source reservoir to the QD $\Gamma_{\rm in}(x)$, which depends on the QD-reservoir distance. This regime can be achieved by starting sufficiently far away from the reservoir such that $\Gamma_{\rm in}$ is much smaller than our measurement resolution of around $50$~fA. Notably, we verified this regime of operation by tip-height dependent measurements closer to the reservoir (supp. mat. Figure~\ref{supp3}). 

Between $x$ = $4.0$ -- $13$~nm where we measure resonant tunnelling only between the reservoir and the QD, we obtain a best fit value for $\Delta I(x)\propto \exp(\kappa x)$ of $\kappa$ = $0.2288$~nm$^{-1}$ (Figure~\ref{fig3}, black line). To relate this to the decay of the induced QD wavefunction in space, we compare the decay to a model for the tunnel coupling $t= \langle \psi_{QD}|V_{QD}|\psi_{res} \rangle$ from the source reservoir to the QD that approximates the tail of the QD wavefunction as $\psi_{QD}\approx A\exp(-|x|/\lambda)$, where $\lambda$ is the decay length. With an abrupt model for the reservoir electrons $\psi_{res}$ (supp. mat.), we obtain $I \propto t^2 \propto \exp(-2|x_{0}|/\lambda)$, where $x_{0}$ is the position of the reservoir relative to the QD, and from this $\lambda=2/\kappa \approx 9$~nm using the extracted $\kappa$. We note that our assumption of an abrupt reservoir makes this an upper bound on the actual decay.  Notably, the slowly varying tail is convenient to controllably tunnel couple the QD to elements such as individual donors in devices. For example, highly tunable tunnel coupling would be advantageous for local spin readout on small scale dopant-based quantum simulators\citep{Salfi2016, Le2017,Le2019} where STM tips can be positioned with sub-nanometre accuracy.

By sweeping the gate voltage $V_{g}$ applied to the gate donor reservoir, we demonstrate the ability to electrically tune the energy of the QD at a fixed tip-sample voltage using the gate. Figure~\ref{fig4}a shows a schematic of our device: capacitive interactions between the QD and the tip, source and gate reservoirs are represented by $C_{t}$, $C_{s}$ and $C_{g}$ respectively, and are present as we keep the tip fixed in space whilst varying $V_{g}$. We vary the gate voltage $V_{g}$ over a $1$~V range from $-2$~V to $-1$~V to examine the gating characteristics of the QD, with the tip held at the position indicated by the marker on the line in the STM inset of Figure~\ref{fig4}b. Figure~\ref{fig4}b shows the conductance through the QD as a function of $V_{s}$ and  $V_{g}$. The green line indicates linear behaviour of the gate during resonance for $V_g\lesssim -1.6$~V. When $V_{g}$ is increased toward $-1$~V, higher values of $V_{s}$ are required to keep the QD state on resonance with the source reservoir. As expected, a more positive gate with an increasingly attractive potential pulls the level of the QD lower in energy. Thus, $V_{s}$ increases and the amount of tip-induced band bending required is lessened, with a corresponding decrease in current through the QD as a consequence of the widening tunnel barrier. 

We analyze the capacitive coupling of the quantum dot, and extract the gate lever arm $\alpha_g$ in the voltage range $V_{g} = -1.8$ to $-1.6$~V where the gate acts linearly. Here, the change in bias voltage required to maintain resonant tunnelling is given by $\partial V_s/\partial V_g = C_g/(C_{\Sigma} - C_s)$, where $C_{\Sigma}$ is the sum of all capacitances between the QD and its environment, including a stray capacitance not shown in Figure.~\ref{fig4}a. In this regime, $\partial V_s/ \partial V_g\sim 0.77$ is extracted from the gradient of the green line. Using  the lever arm extracted from the source bias dependent measurement, where $\alpha_{0} = (C_\Sigma - C_s)/C_\Sigma \sim 0.1$ (supp. mat.), we find that $C_s/C_\Sigma\sim 0.9$ and obtain a gate lever arm for the induced quantum dot of $\alpha_g \equiv C_g/C_\Sigma \sim 0.08$. 
Evidently, the QD capacitive coupling is dominated by $C_s$, with an appreciable gate capacitive coupling, where $C_g/C_t \sim 3.4$ (supp. mat.) despite the metal tip being only $\sim 1$~nm away from the QD. The effect of the gate weakens for values $V_{g} = -1.5$~V and higher and $V_{s}$ flattens out around $\sim -1.215$~V. This behaviour could be due to an enhancement of stray capacitance due to an accumulation of charges in or near the source, by the gate. The control of the induced QD by the gate is important for the potential use of the QD to understand the operation of devices or to probe quantum states induced in devices.

In conclusion, we have demonstrated the ability to induce a single localized electron QD within a device, and to tune its tunnel coupling to a source electrode and electric interaction with a gate electrode within a silicon electronic device. We find that the single electron QD reacts to charges present in the device, both due to individual defects, and due to voltages applied to gates in the device. With the sub-nanometre precision of the tip and the slowly-varying nature of the ground state QD wavefunction, a high level of tunnel coupling control to donor-based electrodes is possible. This opens up the possibility to use tip-induced quantum dots to characterize devices and the quantum states they host, and even to add new device functionality. We expect this technique will be applicable to advanced devices based on materials that can be engineered at the atomic scale, because it is compatible with LT-STM, the most common technique for probing atomic-scale materials and assembling atomic-scale devices\citep{Folsch2014,Huff2018}.

\section*{Acknowledgements}
K.S.H.N. would like to acknowledge S. Loth for useful discussions. We acknowledge support from the ARC Centre of Excellence for Quantum Computation and Communication Technology (CE170100012) and an ARC Discovery Project (DP180102620). J.S. acknowledges financial support from an ARC DECRA fellowship (DE1601101490) and from the National Science and Engineering Research Council. B.C.J. and J.C.MC acknowledge the AFAiiR node of the NCRIS Heavy Ion Capability for access to ion-implantation facilities.

\section*{Author Contributions}
J.S. proposed the idea to probe an electronic device using an induced quantum dot, with input from B.V. and S.R. S.R. proposed the in-situ devices scheme with input from B.V. and J.S. K.S.H.N. performed all the measurements and analysis with input from J.S., B.V. and S.R. B.V., B.C.J. and J.C.MC. led the design, fabrication, vacuum preparation and measurement scheme for in-situ use of the device with input from J.S. and S.R. K.S.H.N. wrote the manuscript with guidance from J.S. and input from B.V. and S.R.

\clearpage
\section*{Supplementary Information}
\subsection{Device Fabrication}
All measurements are performed using an Omicron LT STM at $4.2$~K. A commercial p-type boron doped silicon wafer is used with a doping density of $\approx 10^{15-16}$~cm$^{-3}$, corresponding to a resistivity of $\approx $1--10 $\Omega\cdot$~cm. Antimony ion beam implantation is performed such that a dopant layer is created just below the silicon surface with a density of $\approx $ 1$\times$10$^{15}$~cm$^{-2}$ at selected regions of the sample with the use of a $120$~nm thick HSQ oxide mask. Once cleaned, the sample is degassed at $600~\degree$C for 12 hours. Following this, two additional flashes at $\approx$ $1000~\degree$C are performed, each for 10 seconds. The temperature is then rapidly brought to $800~\degree$C before slowly decreasing the temperature at a rate of $10~\degree$C/second to $340~\degree$C to allow for the silicon surface reconstruction. Hydrogen passivation is then performed at $340~\degree$C for approximately 6 minutes at a pressure of $6\times10^{-7}$~mbar before transferring the sample to the STM for measurements with a base temperature of $2\times10^{-11}$~mbar.

\subsection{Additional resonance and gating characteristics}
\begin{figure*}[h]
\includegraphics{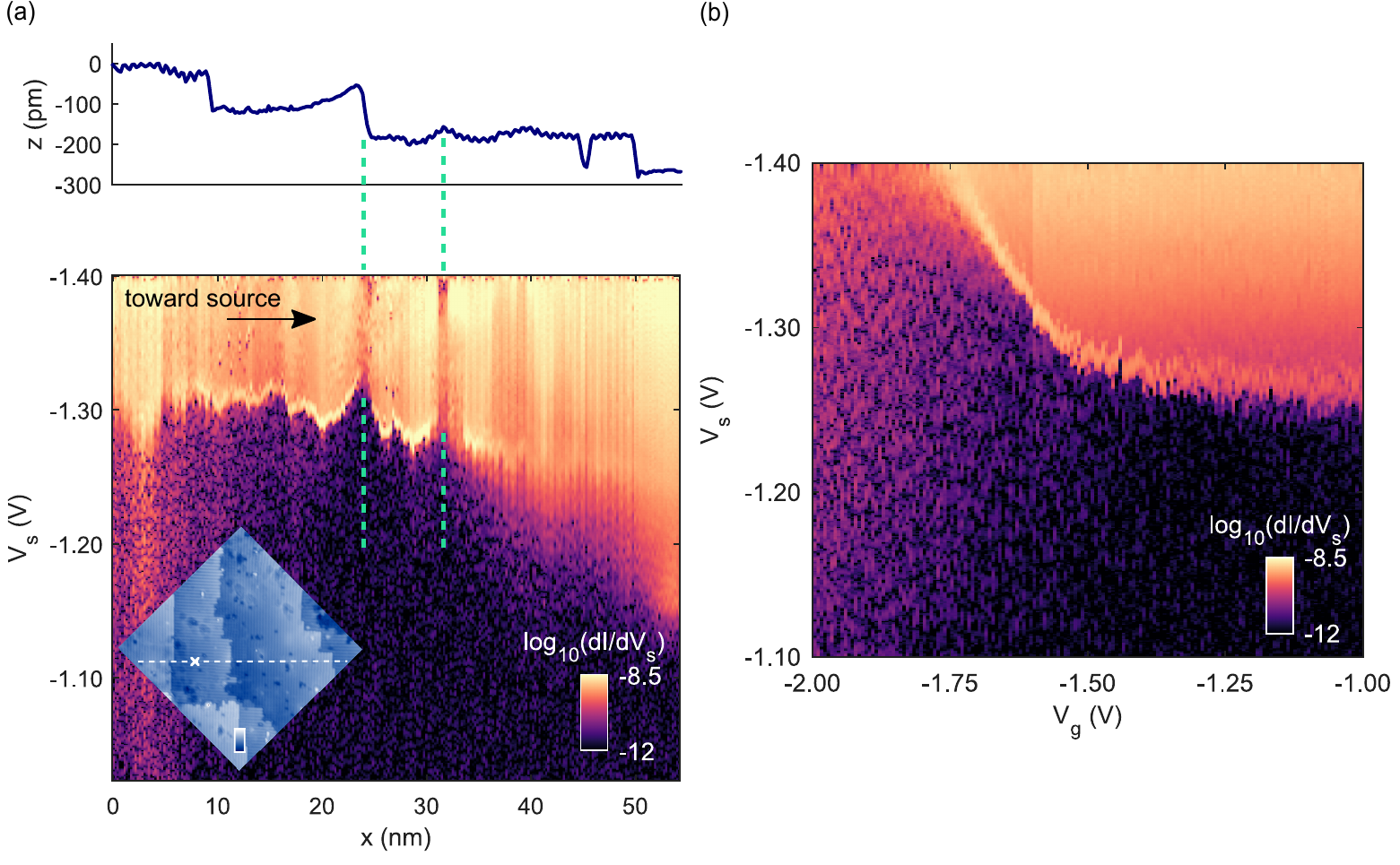}
		\caption[resonance_2]
		{(a) Spectroscopy measurements along a line at a different position in $y$ traversing the same undoped-source junction, $58$~nm away from the measurements of Figure~\ref{fig2}a. A resonance is again observed with characteristics of the same tip-induced QD in Figure~\ref{fig2}a. As the QD travels along the line indicated in the STM image inset, the tunnelling resonance is disturbed by the QD encountering a step edge and a dangling bond indicated by the green dashed lines at $x$ = $24$~nm and $32$~nm respectively. From the topography above, the tip moves up $60$~pm in $z$ as it approaches the step edge of the terrace between $x$ = $10$ -- $23$~nm, likely due to accumulated negative charge at the step edge. At $x$ = $32$~nm, the small rise in topography reflects the tip passing over the edge of a negatively charged (2e$^{-}$) dangling bond, where a bright ring surrounds the characteristic halo \citep{Labidi2015} of a dangling bond that is momentarily positive when imaged by the tip in an un/weakly doped sample. Both sources of localized charge increase the QD energy before disturbing resonant tunnelling, as expected for a repulsive interaction. As the tip continues to approach the reservoir, the resonance is eventually not visible, indicating that the QD has been fully driven in to a region of the reservoir with a high donor density.  Resonant tunnelling is no longer observed and the tip then probes the filled state density of the donor reservoir. Taken at $V_{g}$ = $-1.6$~V. Inset: topography image taken at $V_{s}$ = $-1.6$~V. Scale: $0$ -- $0.36$~nm. (b) Gating characteristics of the resonance taken at the position indicated by the marker on the line of the inset in (a). The same gating behaviour is seen for the QD when compared to Figure~\ref{fig4}b.}\label{supp1}
\end{figure*}

\begin{figure*}[h]
\includegraphics{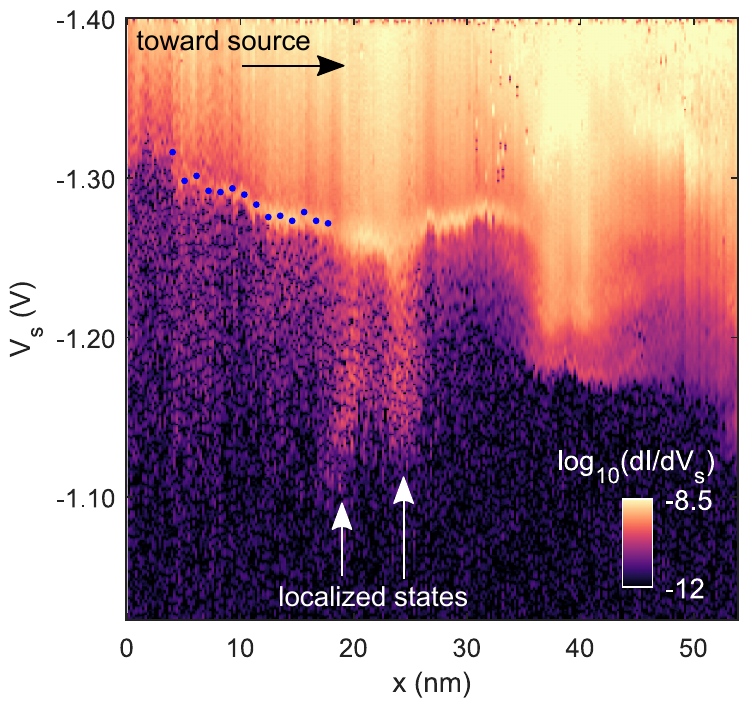}
		\caption[resonance_extended]
		{A separate and extended spectroscopy measurement along the same line used for the measurements of Figure~\ref{fig2}. The QD is driven deep in to the donor reservoir, indicated by the resonance not observed for $x\gtrsim 35$~nm. By moving the tip and QD beyond $x$ = $18$~nm, we observe the QD interacting with localized states around $x \approx 17$ -- $25$~nm and also at $x\approx 35$ -- $50$~nm for voltages between $-1.25$~V to $-1.12$~V, which requires less tip-induced band-bending compared to the QD state. The blue dots plotted on the resonance show the conductance peak positions used to calculate $\Delta I(x)$. Taken at $V_g = -1.6$~V.}\label{supp2}
\end{figure*}

\begin{figure*}[h]
\includegraphics{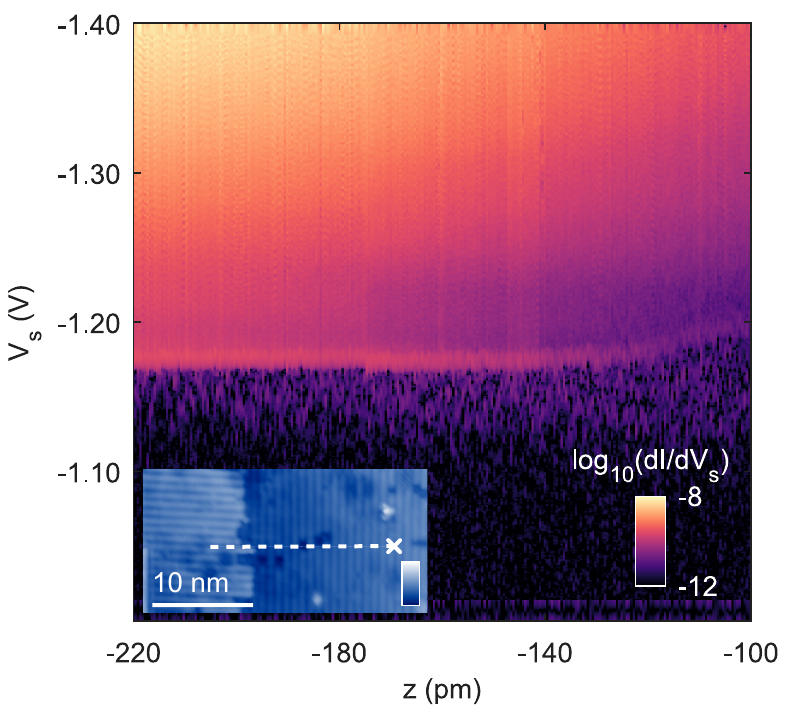}
		\caption[tip_height_dependence]
		{Conductance map during resonant tunnelling as a function of tip height $z$, with the tip placed at the position of the marker on the line of the STM inset ($x = 18$~nm). Despite the slight reduction in conductance for $x\gtrsim 13$~nm, $\Gamma_{\rm in}$ here at $x = 18$~nm is larger than its corresponding value when the resonance is first observed at $x \approx 4$~nm. At this point where the tip is closest to the reservoir (for Figure~\ref{fig2}), no significant change in the conductance peak height and lineshape is observed $\pm$ $20$~pm from $z = -200$~pm, the tip height setpoint used for the measurement of Figure~\ref{fig2}. Thus, the tunnelling regime $\Gamma_{\rm in} \ll \Gamma_{\rm out}$ is established during resonant tunnelling. Done when $V_{g}$ = $-1.4$~V, a $0.2$~V difference in $V_{g}$ is not expected to result in any change in the tunnelling regime. As $z$ is increased, the resonance weakens and is eventually not visible as expected. Inset scale: $0$ -- $0.38$~nm.}\label{supp3}
\end{figure*}

Additional spectroscopy measurements discussed in the main text are presented in Figure~\ref{supp1} and \ref{supp2}.

\clearpage
\subsection{Calculation of \textit{t} dependence on $\psi_{QD}$}

\begin{figure} [h]
\includegraphics{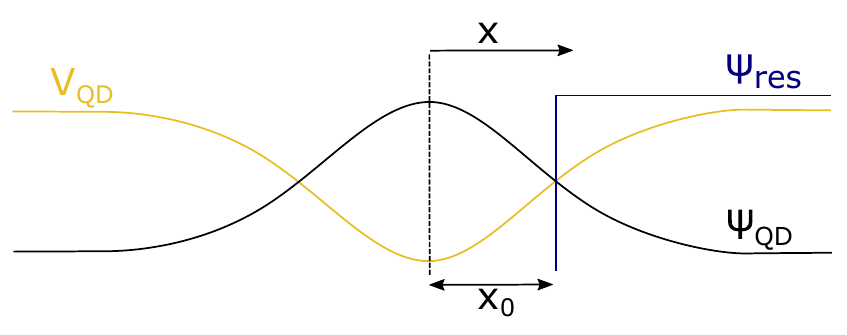}
		\caption[t schematic]
		{Simplified schematic model of reservoir/QD wavefunction overlap.}\label{supp4}
\end{figure}

Assuming the QD potential $V_{QD}$ is parabolic around the tip ($\sim$ QD centre)\citep{Dombrowski1999} while slowly varying in $x$ and $y$ to 0 far away from the tip, we approximate the tail of $\psi_{QD}$ with an exponentially decaying function $\psi_{QD} = A\exp(-|x|/ \lambda)$ and assume the donor reservoir $\psi_{res}$ as homogeneous, normalized to 1 for $x > x_{0}$, the distance between QD centre and reservoir edge, and 0 otherwise. This is schematically illustrated in Figure~\ref{supp4}.

\begin{align*}
  t
  &= \langle \psi_{QD}|V_{QD}|\psi_{res} \rangle\\
  &= AV_{QD}\int_{-\infty}^{\infty} \exp(-|x|/ \lambda) \psi_{res} dx\\
  &= AV_{QD}\int_{x_{0}}^{\infty} \exp(-|x|/ \lambda) dx \\
  &= -AV_{QD}\lambda \left[\exp(-|x|/ \lambda)\right]_{x_{0}}^{\infty}\\
  &= AV_{QD}\lambda \exp(-|x_{0}|/ \lambda)
\end{align*}
 
$I \sim t^{2} \sim \exp(-2|x_{0}|/ \lambda) $ as $\exp(-2|x_{0}|/ \lambda)$ is faster than $\lambda^2$. So as the QD approaches the reservoir i.e. $x$ increases, $x_{0}$ decreases and the current $I$ increases exponentially as measured.

\clearpage
\subsection{Calculation of the gate lever arm and additional capacitance ratios}

\begin{figure} [h]
\includegraphics{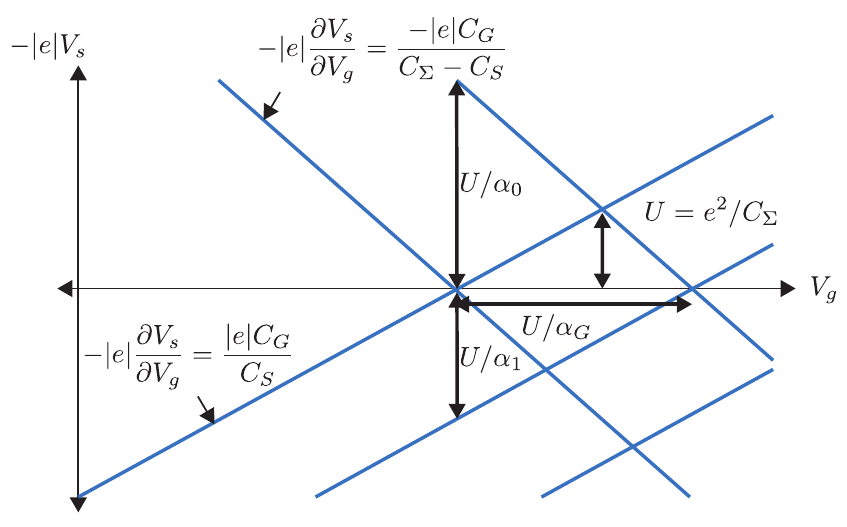}
		\caption[t schematic]
		{Stability diagram of a single-electron transistor. One of the measured slopes and a measured lever arm parameter are used together to deduce the gate lever arm $\alpha_G$, which could not be deduced directly from our experiment. }\label{supp5}
\end{figure}

The two measured quantities $\partial V_{s}/\partial V_{g}$ and the bias lever arm $\alpha_0$ can be deduced using only slopes from the Coulomb diagram pattern of the orthodox model for single-electron tunneling. This model is defined in terms of the source capacitance $C_S$, the gate capacitance $C_G$, the tip capacitance $C_T$, and a stray capacitance $C_0$ (see Fig. ~\ref{fig4}a), and the total QD self-capacitance $C_\Sigma = C_S + C_G + C_T + C_0$, and the stability diagram is shown in Fig.~\ref{supp5} from reference \onlinecite{Hanson2007}.

First, from Fig.~\ref{supp5} we can deduce that for $V_S<0$, the measured parameter $\partial V_{s}/\partial V_{g}$ corresponds to $C_G/(C_\Sigma-C_S)$, which was experimentally determined to be $0.77$.  Second, from Fig.~\ref{supp5} we can also deduce that $\alpha_0 = (C_\Sigma - C_S)/C_\Sigma$, which was experimentally determined to be $\alpha_0 \approx 0.1$, and $\alpha_1=C_S/C_\Sigma$.  We can re-write the definition of the gate lever arm $\alpha_G\equiv C_G/C_\Sigma$ in terms of measured quantities $\alpha_0$ and $\partial V_{s}/\partial V_{g}$ as follows: 

\begin{equation}
\alpha_G\equiv \frac{C_G}{C_\Sigma} = \frac{C_G}{C_\Sigma - C_S} \frac{C_\Sigma - C_S}{C_\Sigma}\nonumber
\end{equation}

which is equivalent to $\alpha_g = \alpha_0 \partial V_{s}/\partial V_{g}\approx 0.1 \times  0.77 \approx 0.08$ as discussed in the main text. 
\\
\\
Additional capacitance ratios comparing the coupling between the QD and source, gate and tip are calculated (assuming negligible stray capacitance $C_0$) as follows:

\begin{equation}
\frac{C_G}{C_T} = \frac{C_G/(C_G + C_T)}{1-C_G/(C_G + C_T)} = \frac{0.77}{1-0.77} \approx 3.4\nonumber
\end{equation}

\begin{equation}
\frac{C_S}{C_G} = \frac{1-\alpha_0}{\alpha_0 C_G/(C_G + C_T)} =  \frac{1-0.1}{0.1\times 0.77} \approx 12\nonumber
\end{equation}

\begin{equation}
\frac{C_S}{C_T} = \frac{C_S}{C_G} \frac{C_G}{C_T} \approx  39\nonumber
\end{equation}

\clearpage
\bibliography{bibliography}

\end{document}